\documentclass[10pt,conference]{IEEEtran}
\IEEEoverridecommandlockouts

\newcommand{\IEEEAcceptedNotice}{%
To appear in the Proceedings of IEEE CASCON 2026.\\
\copyright~2026 IEEE. Personal use of this material is permitted.  Permission from IEEE must be obtained for all other uses, in any current or future media, including reprinting/republishing this material for advertising or promotional purposes, creating new collective works, for resale or redistribution to servers or lists, or reuse of any copyrighted component of this work in other works.
}

\makeatletter
\def\ps@IEEEtitlepagestyle{%
  \def\@oddfoot{%
    \parbox[b]{0.96\textwidth}{%
      \centering
      \scriptsize
      \IEEEAcceptedNotice
    }%
  }%
  \def\@evenfoot{}%
}
\makeatother
\usepackage{cite}
\usepackage{amsmath,amssymb,amsfonts}
\usepackage{algorithmic}
\usepackage{graphicx}
\usepackage{textcomp}
\usepackage{hyperref}
\usepackage{xcolor}
\def\BibTeX{{\rm B\kern-.05em{\sc i\kern-.025em b}\kern-.08em
    T\kern-.1667em\lower.7ex\hbox{E}\kern-.125emX}}
\begin{document}

\title{Online AutoML: Evaluating Poisoning Attacks on Cost-Utility Adversarial Training Defence Strategy for IoT Networks}


\author{
    \IEEEauthorblockN{
        \IEEEauthorrefmark{1}\IEEEauthorrefmark{2}Chukwunonso Henry Nwokoye,
        \IEEEauthorrefmark{1}Khalil El-Khatib,
        and \IEEEauthorrefmark{1}Li Yang
    }
    \IEEEauthorblockA{
        \IEEEauthorrefmark{1}\textit{Faculty of Business and Information Technology,}
    \textit{Ontario Tech University,}
        Oshawa, Ontario, Canada \\
        \IEEEauthorrefmark{2}\textit{Department of Computer Science,}
    \textit{Alex Ekwueme Federal University,}
        Nigeria \\
        Emails: \{henrychukwunonso.nwokoye, khalil.el-khatib, li.yang\}@ontariotechu.ca
    }
}

\maketitle

\begin{abstract}
Machine learning (ML)-powered poisoning attack vectors are adversarial maneuvers whereby an attacker intentionally inserts, corrupts, or alters training data to distort an ML model's learning process. The objective is to diminish model efficacy, instill biases, induce misclassifications, or include concealed backdoors that may be attacked during implementation. In streaming contexts, poisoning attacks pose significant risks since models perpetually update based on incoming streams of data. An assailant may incrementally introduce harmful samples into this data stream, leading the model to assimilate erroneous features over time without timely identification. Therefore, this study is aimed at evaluating the efficacy of the adversarial training (AT) defense approach against poisoning attacks (label flip and noise injection) using an online AutoML pipeline for Internet of Things (IoT) networks. Specifically, poisoning attacks (label flip and noise injection) were applied to streaming-capable AutoML learners (Hoeffding Tree (HT), Leveraging Bagging (LB), Adaptive Random Forest (ARF), Hoeffding Adaptive Tree (HAT), and Streaming Random Patches (SRP)). Under the strongest poisoning rate (PR = 1.0), AT-SRP achieved the highest F1-score against label flip poisoning (0.904), while AT-LB achieved the highest F1-score against noise-injection poisoning (0.933). Finally, several drift detection methods were used for rolling accuracy and prequential evaluation. 

\end{abstract}

\begin{IEEEkeywords}
Internet of Things (IoT), Security, Online AutoML, Data Streams, Automated Data Analytics, Concept Drift, Online Machine Learning
\end{IEEEkeywords}

\section{Introduction} 


Online machine learning refers to a family of algorithms that learn incrementally from data arriving as a continuous stream, updating their models with each new instance rather than retraining on static datasets \cite{gomes2019machine}. This paradigm is fundamentally suited to environments where data volumes are unbounded and computational resources are constrained, requiring models to process each instance within bounded time and memory \cite{ziffer2023towards}. Data streams are characterised by high velocity, potentially infinite length, and non-stationarity, a property known as concept drift, in which the underlying data distribution shifts over time \cite{montiel2021river}. Stream classifiers must therefore adapt continuously, remaining responsive to distributional changes without storing or revisiting prior instances \cite{gomes2019machine}.

The practical relevance of online learning extends across several security-critical domains. In cybersecurity, the volume and velocity of network traffic make batch retraining impractical, necessitating adaptive models that classify data in real time \cite{ceschin2024machine}. Intrusion Detection Systems (IDS) deployed in IoT networks face compounded challenges: the heterogeneous nature of IoT devices, their resource constraints, and the diversity of attack vectors demand classifiers that learn continuously from evolving traffic \cite{hnamte2023dcnnbilstm}. Online AutoML frameworks, such as the River library formed through the merger of creme and scikit-multiflow, automate the selection and configuration of the best-performing stream learner within a unified pipeline, making them particularly attractive for IoT deployments \cite{montiel2021river}. Beyond IoT security, online learners are central to financial fraud detection, where transaction streams must be classified within milliseconds against continuously shifting fraud patterns \cite{Carcillo2017}.

Despite the growing adoption of online AutoML systems in security-sensitive applications, their resilience to training-time adversarial manipulation remains poorly understood \cite{ceschin2024machine}. Poisoning attacks in which an adversary introduces corrupted instances into the training stream to degrade model performance pose a direct and realistic threat to continuously learning systems \cite{tian2022comprehensive}. Unlike static batch models, stream classifiers update with every incoming instance, meaning each poisoned data point immediately distorts the learned decision boundary \cite{biggio2018wild}. Label-flipping attacks, a widespread form of poisoning, deliberately invert the ground-truth labels of selected training instances to steer the classifier toward systematic misclassification of targeted classes \cite{aryal2022analysis}. This vulnerability is further amplified by drift detection mechanisms embedded within online pipelines: a detector influenced by adversarially induced error patterns may trigger spurious drift alarms, causing premature model resets and destabilising the learning process \cite{sethi2017reliable}.

Building on the above studies, this paper makes three primary contributions. First, we develop a poisoning attack framework targeting the training stream of an online AutoML pipeline, implementing label flip and noise injection \cite{chan2018algorithm} on the whole feature space at varying poisoning rates 
\footnote{Code and supplementary results are available at: \url{https://github.com/ChiNonsoHenry16/Label-Flip-NoiseInjection_OnAutoML}}.

Second, we evaluate the degradation in accuracy, precision, recall, and F1-score across five stream classifiers: Hoeffding Tree (HT), LeveragingBagging (LB), Adaptive Random Forest (ARF), Hoeffding Adaptive Tree (HAT), and Streaming Random Patches (SRP). Third, we analyze the behavioral response of the Early Drift Detection Method (EDDM), Adaptive Windowing (ADWIN), and Drift Detection Method (DDM) drift detectors under poisoned conditions and evaluate the adversarial training defense. Together, these contributions address an identified gap in the security of online AutoML systems for IoT environments \cite{ceschin2024machine, hallaji2023label}.

\section{RELATED WORK}
\label{lit_rew}


Poisoning attacks constitute a class of training-time adversarial threats in which an attacker manipulates the data used to build a machine learning model either to degrade overall accuracy (availability attacks) or to induce targeted misclassifications (integrity attacks) \cite{biggio2018wild}. Foundational work by Biggio et al. \cite{biggio2012poisoning} established that support vector machines are susceptible to gradient-based poisoning, providing the theoretical basis for subsequent research in this area. Label-flipping attacks corrupt training labels without modifying feature values, making them particularly difficult to detect through conventional input validation \cite{taheri2020defending}. These attacks are notoriously hard to detect at scale. Training datasets are often too large for per-instance inspection, and adversarial label flips can be subtle, especially in continuously updating systems \cite{yerlikaya2022data}. 

Recent empirical evidence confirms that even modest poisoning rates as low as two to five percent of the training distribution are sufficient to cause significant classification degradation  \cite{aryal2022analysis}. In IoT contexts, Tian et al. \cite{tian2022comprehensive} conducted a comprehensive survey of poisoning strategies and countermeasures, observing that edge-deployed systems relying on live data streams are particularly vulnerable due to the absence of offline validation checkpoints. The attack surface widens considerably in federated and distributed IoT learning scenarios, where model updates from multiple heterogeneous devices are aggregated at a central server \cite{sun2021data}. Nguyen et al. \cite{nguyen2020poisoning} demonstrated that federated IDS systems can be severely compromised by even a small number of malicious participants injecting poisoned local updates into the aggregation process. Yang et al. \cite{yang2023clean} examined clean-label poisoning in federated IoT environments, showing that adversaries can embed malicious behaviour within feature-plausible samples, effectively evading statistical anomaly detectors. In a clean label attack, the attacker introduces alterations to the features while maintaining the labels in an ostensibly "clean" state. This is extensively researched and particularly relevant in situations when an assailant is unable to alter labels (for instance, automated or crowdsourced data management). Noise injection can be classified as clean-label and was also specifically characterized as a data poisoning strategy \cite{chan2018algorithm}, whereby hostile noise injected into training data distorts deep learning frameworks in a visually undetectable way. Collectively, these findings confirm that poisoning represents an operationally viable threat to deployed IoT machine learning pipelines, not merely a theoretical concern \cite{tian2022comprehensive}.

Despite substantial progress in adversarial robustness for batch machine learning, the security of online AutoML pipelines against poisoning remains sparsely studied \cite{ceschin2024machine}. AT offers a more practical defense posture for resource-constrained online learners; however, its systematic evaluation against stream classifiers in IoT security settings has not been conducted \cite{hallaji2023label}. Likewise, while adversarial concept drift detection has been explored in isolation \cite{sethi2017reliable}, its behavior within a full online AutoML pipeline comprising multiple learner options has not been studied. This paper directly addresses these gaps by providing the first joint evaluation of poisoning attack impact and adversarial defense across HT, LB, ARF, HAT, and SRP in a unified streaming AutoML setting.

AT has been named among defense strategies against poisoning attacks \cite{ALLHEEIB2024}. Therefore, the AT approach  was adapted along with the dirty (label flip) and clean label \cite{yang2023clean} (noise injection) \cite{chan2018algorithm} poisoning attacks. Considering the papers reviewed herein, we aim to extend the online AutoML pipeline proposed by Yang \& Shami  \cite{yang2022iot} by applying the AT \cite{kireev2022adversarial}. Furthermore, the impact of poisoning attacks on the improved pipeline was evaluated using popular metrics.  Also, we implemented live learning to manage concept/model drift in real-time traffic under poisoning attacks.

\section{Methodology}
\label{met}

\begin{figure}
    \centering
    \includegraphics[width=1.0\linewidth]{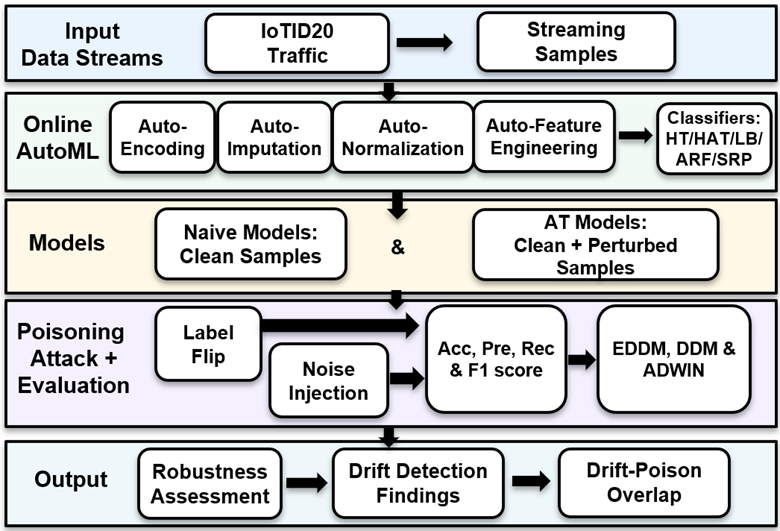}
    \caption{The overview of the proposed framework}
    \label{fig:framework}
\end{figure}

\color{black}
We implement an automated machine learning (AutoML) pipeline, subsequently applying data poisoning assaults in both label and feature domains, and using AT as a defensive strategy, assessed by streaming tree-based classifiers with live concept drift detection. The study employed the IoTID20 dataset \cite{10.1007/978-3-030-47358-7_52}, and performance was measured using accuracy, precision, recall, and F1-score. The proposed methodology is shown in Fig. \ref{fig:framework}. 

Initially, data undergoes automated preprocessing by the identification and encoding of categorical variables via label encoding, then addressing infinite and missing values by substituting them with zeros. Feature normalization was dynamically determined using a Shapiro-Wilk normality assessment: if features exhibit statistical Gaussianity, z-score normalization is applied; if not, min-max scaling transforms features to the range of ([0, 1]). Dimensionality reduction was achieved by prioritizing features based on their relevance derived by LightGBM regression and progressively eliminating those that contribute less than 10\% to the total importance. Superfluous characteristics exhibiting a strong pairwise Pearson correlation $(r > 0.9)$ are then eliminated, retaining just a single feature from each associated cluster.

To assess the resilience to poisoning, two offensive techniques are examined. Label flip (LF) poisoning arbitrarily substitutes the labels of chosen training data with a different class, determined evenly at random. This traditional "dirty-label" assault \cite{biggio2012poisoning, rosenfeld2020certified} instills potentially significant confusion into model training. Noise injection (NI) poisoning, seen as a feature-space and "clean-label" assault \cite{chan2018algorithm}, introduces random uniform (adversarial) noise into each feature of chosen samples while preserving the labels. Each feature value is altered by a stochastic value inside \(\pm 10\%\) of its range and constrained to its domain, emulating feature-space pollution of the input distribution.

AT is used as a defensive strategy for the ensemble and streaming classifiers within River. For each training batch, subsequent to possible contamination, we generate an extra AT sample using the same assault methodology and include both in the model update batch. This resembles the AT framework proposed by Goodfellow et al. \cite{goodfellow2014explaining}, although it operates in a streaming, online context.  Subsequent to training, we calculate the typical classification metrics—accuracy, recall, precision, and F1 score—for both the Naive and AT variants at several poisoning rates (PR). Furthermore, we implement 3 drift detectors (EDDM, DDM, and ADWIN) on the classifier output streams, documenting drift occurrences and examining the relationship between drift instances and the locations of injected anomalies.  For all experiments, we used predetermined random seeds that guarantee reproducibility and facilitate direct scenario comparisons.

The streaming classifiers considered for this study include LB, HAT, ARF, SRP, and HT \cite{yang2022iot}. We primarily selected them to ensure architectural variety among streaming classifiers, including a foundational incremental and drift-responsive tree (HT and HAT) and 3 ensemble techniques (LB, ARF, and SRP). In the experimental setup, we developed naive and adversarially trained (AT) models for each of the online learners (HT, LB, ARF, SRP, and HAT). The resulting models are Naive-HT, AT-HT, Naive-LB, AT-LB, Naive-SRP, AT-SRP, Naive-ARF, AT-ARF, Naive-HAT, and AT-HAT. Implementing both naive and AT model variants is important for assessing robustness against poisoning attacks \cite{Yang2024}.  



Poisoning attacks affect the data during the training phase. In this scenario, the assailant introduces harmful or adversarial instances into the training dataset to distort or skew the model throughout its training period (in some cases, a portion of the labels may be altered (label flipping), attributes modified, etc.). Besides label flip, we executed a feature-space data contamination assault by including adversarial noise into training samples \cite{chan2018algorithm}, whereby noise-infused data is used to intentionally corrupt the learning process. 


\section{Results}
\label{res}

\color{black}
In the evaluation of the proposed methodology for adversarial ML threats and control, it is essential to report metrics for naive and AT model variants. This is usually referred to as adversarial robustness research \cite{madry2017towards}, and it allows security managers to understand the extent to which a particular defense strategy impacts both adversarial attacks and the metrics. 


Here, we present poisoning rate (PR) as well as the Naive and AT variants of metrics such as accuracy, precision, recall and F1 score. The PR represents the proportion (\%) of the training dataset that is subjected to manipulation (modification). It indicates the quantity of contaminated training instances. Note that PR ranges from 0.00 to 1.00. To succinctly illustrate the effects of the poisoning attacks (label flip and noise injection) at different intensities, we present findings for three illustrative conditions, namely (a) No attack/baseline (PR = 0.00), (b) mild attack (PR = 0.40), and (c) strong attack (PR = 1.00). These represent the baseline, a plausible adversarial situation, as well as an upper-bound worst-case circumstance, respectively. For every online classifier, we compared performances under these conditions to distinctly demonstrate the susceptibility of the naïve model and the resilience afforded by the AT approach. The comprehensive results are available at the following GitHub repository: \url{https://github.com/ChiNonsoHenry16/Label-Flip-NoiseInjection_OnAutoML}.

\subsection{Impact of Poisoning Attacks}

The label-flip outcomes reveal a significant disparity between the naïve and AT setups (Tables \ref{tab:naive_label_flip} and \ref{tab:at_label_flip}). In the naive setup, raising the PR often results in significant deterioration, with several classifiers attaining F1-scores around zero at a PR of 1.0. Conversely, AT offers significantly enhanced robustness for some models. Significantly, SRPClassifier attains an F1-score of 0.904, an accuracy of 0.906, and a recall of 0.906 at the 1.0 label-flip rate, while HAT records an F1-score of 0.778. Utilizing LB maintains an F1 score of 0.661 at the maximum PR of 1.00.

\begin{table}[!t]
\caption{Naive Model Performance Under Label-Flip Poisoning}
\label{tab:naive_label_flip}
\centering
\resizebox{\columnwidth}{!}{%
\begin{tabular}{llcccc}
\hline
\textbf{Classifier} & \textbf{Rate} & \textbf{Accuracy} & \textbf{F1} & \textbf{Precision} & \textbf{Recall} \\
\hline
Hoeffding Tree & 0.0 & 0.942 & 0.914 & 0.888 & 0.942 \\
Hoeffding Tree & 0.4 & 0.942 & 0.914 & 0.888 & 0.942 \\
Hoeffding Tree & 1.0 & 0.058 & 0.006 & 0.003 & 0.058 \\
\hline
Leveraging Bagging & 0.0 & 0.898 & 0.911 & 0.928 & 0.898 \\
Leveraging Bagging & 0.4 & 0.191 & 0.259 & 0.857 & 0.191 \\
Leveraging Bagging & 1.0 & 0.058 & 0.006 & 0.003 & 0.058 \\
\hline
SRPClassifier & 0.0 & 0.936 & 0.911 & 0.888 & 0.936 \\
SRPClassifier & 0.4 & 0.286 & 0.391 & 0.880 & 0.286 \\
SRPClassifier & 1.0 & 0.055 & 0.049 & 0.452 & 0.055 \\
\hline
Hoeffding Adaptive Tree & 0.0 & 0.942 & 0.914 & 0.888 & 0.942 \\
Hoeffding Adaptive Tree & 0.4 & 0.942 & 0.914 & 0.888 & 0.942 \\
Hoeffding Adaptive Tree & 1.0 & 0.058 & 0.006 & 0.003 & 0.058 \\
\hline
Adaptive Random Forest & 0.0 & 0.942 & 0.914 & 0.888 & 0.942 \\
Adaptive Random Forest & 0.4 & 0.839 & 0.860 & 0.882 & 0.839 \\
Adaptive Random Forest & 1.0 & 0.058 & 0.006 & 0.003 & 0.058 \\
\hline
\end{tabular}%
}
\end{table}

\begin{table}[!t]
\caption{AT Model Performance Under Label-Flip Poisoning}
\label{tab:at_label_flip}
\centering
\resizebox{\columnwidth}{!}{%
\begin{tabular}{llcccc}
\hline
\textbf{Classifier} & \textbf{Rate} & \textbf{Accuracy} & \textbf{F1} & \textbf{Precision} & \textbf{Recall} \\
\hline
Hoeffding Tree & 0.0 & 0.058 & 0.006 & 0.003 & 0.058 \\
Hoeffding Tree & 0.4 & 0.058 & 0.006 & 0.003 & 0.058 \\
Hoeffding Tree & 1.0 & 0.058 & 0.006 & 0.003 & 0.058 \\
\hline
Leveraging Bagging & 0.0 & 0.429 & 0.550 & 0.905 & 0.429 \\
Leveraging Bagging & 0.4 & 0.296 & 0.409 & 0.860 & 0.296 \\
Leveraging Bagging & 1.0 & 0.543 & 0.661 & 0.871 & 0.543 \\
\hline
SRPClassifier & 0.0 & 0.112 & 0.124 & 0.830 & 0.112 \\
SRPClassifier & 0.4 & 0.153 & 0.191 & 0.880 & 0.153 \\
SRPClassifier & 1.0 & \textbf{0.906} & \textbf{0.904} & 0.902 & \textbf{0.906} \\
\hline
Hoeffding Adaptive Tree & 0.0 & 0.942 & 0.914 & 0.888 & 0.942 \\
Hoeffding Adaptive Tree & 0.4 & 0.058 & 0.006 & 0.003 & 0.058 \\
Hoeffding Adaptive Tree & 1.0 & 0.703 & 0.778 & 0.873 & 0.703 \\
\hline
Adaptive Random Forest & 0.0 & 0.123 & 0.133 & 0.909 & 0.123 \\
Adaptive Random Forest & 0.4 & 0.076 & 0.064 & 0.673 & 0.076 \\
Adaptive Random Forest & 1.0 & 0.058 & 0.006 & 0.003 & 0.058 \\
\hline
\end{tabular}%
}
\end{table}

\begin{table}[!t]
\caption{Naive Model Performance Under Noise-Injection Poisoning}
\label{tab:naive_noise}
\centering
\resizebox{\columnwidth}{!}{%
\begin{tabular}{llcccc}
\hline
\textbf{Classifier} & \textbf{Rate} & \textbf{Accuracy} & \textbf{F1} & \textbf{Precision} & \textbf{Recall} \\
\hline
Hoeffding Tree & 0.0 & 0.942 & 0.914 & 0.888 & 0.942 \\
Hoeffding Tree & 0.4 & 0.905 & 0.922 & 0.951 & 0.905 \\
Hoeffding Tree & 1.0 & 0.942 & 0.914 & 0.888 & 0.942 \\
\hline
Leveraging Bagging & 0.0 & 0.894 & 0.909 & 0.930 & 0.894 \\
Leveraging Bagging & 0.4 & 0.607 & 0.709 & 0.936 & 0.607 \\
Leveraging Bagging & 1.0 & 0.942 & 0.937 & 0.934 & 0.942 \\
\hline
SRPClassifier & 0.0 & 0.942 & 0.914 & 0.888 & 0.942 \\
SRPClassifier & 0.4 & 0.892 & 0.913 & 0.949 & 0.892 \\
SRPClassifier & 1.0 & 0.934 & 0.937 & 0.940 & 0.934 \\
\hline
Hoeffding Adaptive Tree & 0.0 & 0.942 & 0.914 & 0.888 & 0.942 \\
Hoeffding Adaptive Tree & 0.4 & 0.942 & 0.914 & 0.888 & 0.942 \\
Hoeffding Adaptive Tree & 1.0 & 0.942 & 0.915 & 0.888 & 0.942 \\
\hline
Adaptive Random Forest & 0.0 & 0.942 & 0.914 & 0.888 & 0.942 \\
Adaptive Random Forest & 0.4 & \textbf{0.981} & \textbf{0.980} & \textbf{0.980} & \textbf{0.981} \\
Adaptive Random Forest & 1.0 & 0.942 & 0.915 & 0.905 & 0.942 \\
\hline
\end{tabular}%
}
\end{table}

\begin{table}[!t]
\caption{Adversarially Trained Model Performance Under Noise-Injection Poisoning}
\label{tab:at_noise}
\centering
\resizebox{\columnwidth}{!}{%
\begin{tabular}{llcccc}
\hline
\textbf{Classifier} & \textbf{Rate} & \textbf{Accuracy} & \textbf{F1} & \textbf{Precision} & \textbf{Recall} \\
\hline
Hoeffding Tree & 0.0 & 0.686 & 0.769 & 0.907 & 0.686 \\
Hoeffding Tree & 0.4 & 0.718 & 0.791 & 0.905 & 0.718 \\
Hoeffding Tree & 1.0 & 0.721 & 0.793 & 0.905 & 0.721 \\
\hline
Leveraging Bagging & 0.0 & \textbf{0.948} & \textbf{0.952} & 0.958 & \textbf{0.948} \\
Leveraging Bagging & 0.4 & 0.904 & 0.921 & 0.952 & 0.904 \\
Leveraging Bagging & 1.0 & 0.936 & 0.933 & 0.930 & 0.936 \\
\hline
SRPClassifier & 0.0 & 0.923 & 0.917 & 0.912 & 0.923 \\
SRPClassifier & 0.4 & 0.933 & 0.932 & 0.932 & 0.933 \\
SRPClassifier & 1.0 & 0.943 & 0.925 & 0.923 & 0.943 \\
\hline
Hoeffding Adaptive Tree & 0.0 & 0.855 & 0.884 & 0.929 & 0.855 \\
Hoeffding Adaptive Tree & 0.4 & 0.891 & 0.904 & 0.922 & 0.891 \\
Hoeffding Adaptive Tree & 1.0 & 0.803 & 0.848 & 0.910 & 0.803 \\
\hline
Adaptive Random Forest & 0.0 & \textbf{0.946} & \textbf{0.952} & \textbf{0.962} & \textbf{0.946} \\
Adaptive Random Forest & 0.4 & 0.942 & 0.914 & 0.888 & 0.942 \\
Adaptive Random Forest & 1.0 & 0.942 & 0.914 & 0.888 & 0.942 \\
\hline
\end{tabular}%
}
\end{table}

Tables~\ref{tab:naive_noise} and~\ref{tab:at_noise} provide the relevant findings for the noise injection poisoning attack. In contrast to label flipping, noise injection results in a less consistent deterioration across the naïve models, with some classifiers maintaining comparatively good efficacy despite elevated poisoning rates. The AT-based models also have varied reactions. At the peak noise-injection rate, LB attains an F1 score of 0.933, whilst SRPClassifier secures 0.925 and HAT reaches 0.848. Consequently, the findings suggest that AT does not provide consistent enhancements across all classifiers; its efficacy fluctuates based on the poisoning method and the foundational online-learning classifier/framework.

\begin{figure}
    \centering
    \includegraphics[width=1\linewidth]{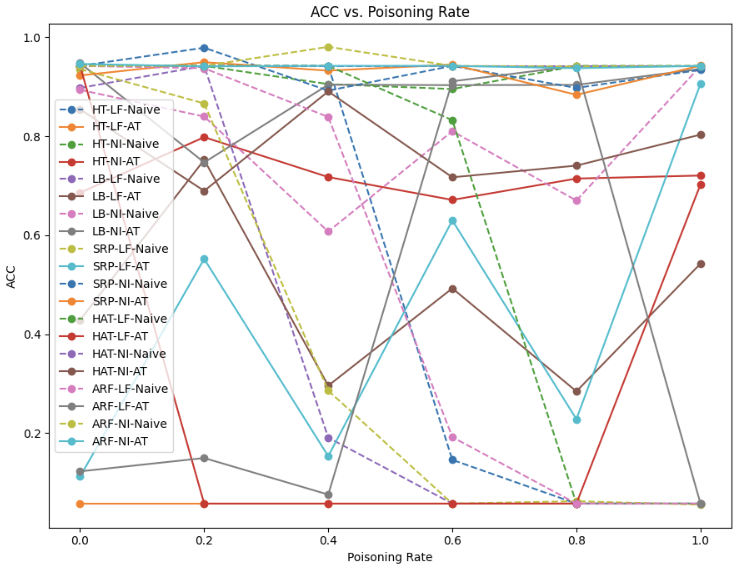}
    \caption{Accuracies and Poisoning Rates}
    \label{fig:PR_Acc}
\end{figure}

The AT findings reveal that SRP classifier is the most robust configuration against label-flip poisoning, whereas the LB is the most effective against noise-injection poisoning, evaluated at the peak non-zero poisoning rate (1.0)  using the F1 score (Table \ref{tab:best_at_models}). In addition, the accuracies and across poisoning rates are depicted in Fig. \ref{fig:PR_Acc}. Graphical illustrations of other metrics available in the GitHub repository provided above.  

\begin{table}[!t]
\caption{Best AT Models Under Poisoning Attacks}
\label{tab:best_at_models}
\centering
\resizebox{\columnwidth}{!}{%
\begin{tabular}{lcccccc}
\hline
\textbf{Poisoning Attack} & \textbf{Best AT Model} & \textbf{Rate} & \textbf{Accuracy} & \textbf{Precision} & \textbf{Recall} & \textbf{F1} \\
\hline
Label Flip 
& \textbf{SRPClassifier} & \textbf{1.0} & \textbf{0.906} & \textbf{0.902} & \textbf{0.906} & \textbf{0.904} \\
Noise Injection 
& \textbf{Leveraging Bagging} & \textbf{1.0} & \textbf{0.936} & \textbf{0.930} & \textbf{0.936} & \textbf{0.933} \\
\hline
\end{tabular}%
}
\end{table}

\subsection{Drift Detection using EDDM}

In this study, we employed several drift detection techniques, including ADWIN and DDM; nevertheless, EDDM revealed a greater number of drift instances for the models.  With EDDM, drift occurrences were rigorously analyzed compared to the others that exhibited lesser sensitivity characteristics for both naïve and AT models under poisoning attacks (label flip and noise injection). Besides highlighting discrepancies in drift detection rates, EDDM's results showed overlap between drift events and the consistency of model classification, at least for noise injection attacks. 

Table~\ref{tab:eddm_drift_results} contains the EDDM detections acquired for the Naive and AT models subjected to LF and NI assaults at poisoning rates of 0.0, 0.4, and 1.0. The findings indicate significant discrepancies in the quantity of identified drifts across different classifiers and kinds of attacks. The Naive-LB model documented 15, 10, and 0 EDDM detections at LF rates of 0.4, 1.0, and 0.0, respectively, but its NI setup yielded 24 detections at a PR of 0.4. The associated AT-LB model yielded 9 and 0 detections at LF rates of 0.4 and 1.0, respectively, and 2 detections with NI at 0.4.

Other classifiers demonstrated varying patterns of EDDM detection. The HT yielded 13 detections for the Naive-HT model at both 0.0 and 0.4 LF rates, whereas AT-HT generated 10 detections across all three LF rates. 

For HAT, the Naive-HAT reliably yielded 13 detections amidst NI at all three rates (0.0, 0.4 and 1.0), while the AT-HAT model recorded 0, 0, and 11 detections, respectively. The ARF exhibited a divergent trend with NI: the Naive-ARF yielded 13, 0, and 4 detections, whereas the AT-ARF model resulted in 3, 13, and 13 detections at rates of 0.0, 0.4, and 1.0, respectively. The findings indicate that EDDM responses fluctuate significantly based on the classifier, the poisoning mechanism and the PR.

\begin{table}[!t]
\caption{EDDM Drift Detection Under Poisoning Attacks}
\label{tab:eddm_drift_results}
\centering
\scriptsize
\begin{tabular}{|l|l|c|c|}
\hline
\textbf{Classifier} & \textbf{Attack} &
\textbf{Naive-EDDM} & \textbf{AT-EDDM} \\
\hline
HoeffdingTree & label\_flip &
13/13/10 & 10/10/10 \\
HoeffdingTree & noise\_injection &
13/0/13 & 6/0/0 \\
\hline
LeveragingBagging & label\_flip &
0/15/10 & 1/9/0 \\
LeveragingBagging & noise\_injection &
0/24/0 & 3/2/0 \\
\hline
SRPClassifier & label\_flip &
2/5/7 & 0/0/2 \\
SRPClassifier & noise\_injection &
13/1/0 & 0/0/0 \\
\hline
HoeffdingAdaptiveTree & label\_flip &
13/13/10 & 13/10/6 \\
HoeffdingAdaptiveTree & noise\_injection &
13/13/13 & 0/0/11 \\
\hline
AdaptiveRandomForest & label\_flip &
13/3/10 & 16/0/10 \\
AdaptiveRandomForest & noise\_injection &
13/0/4 & 3/13/13 \\
\hline
\end{tabular}

\vspace{1mm}
\parbox{\columnwidth}{\scriptsize
\textit{Note:} Each triplet represents the number of EDDM detections at poisoning rates 0.0, 0.4, and 1.0, respectively.
}
\end{table}

\subsection{Drift-Poison Overlap}

To measure the correlation between our proposed poisoning attack models and identified concept drifts, we investigated whether EDDM drift occurrences aligned with samples altered by the poisoning attack, termed drift-poison overlap. Table~\ref{tab:drift-poison} encapsulates the aggregate count of EDDM-identified drifts with the tally of drifts that matched with contaminated samples for chosen noise-injection trials.

No overlap between drift and poison was seen for the label-flip assaults; hence, these findings are omitted from the table. Throughout the label-flip trials, the identified drift occurrences did not align with any contaminated samples. Conversely, little overlap was seen for some NI attacks. For instance, in the case of the LB classifier with a poisoning rate of 0.4, one of the 24 EDDM-identified drifts in the Naive-LB model and one of the two identified drifts in the AT model aligned with a poisoned sample. An overlap of 1 was similarly seen for the AT-SRP and AT-ARF models at the specified NI rates.


\begin{table}[!t]
\caption{Poisoning Attack Results and Drift Detection}
\label{tab:drift-poison}
\centering
\small
\begin{tabular}{|l|c|l|c|c|}
\hline
\textbf{Model} & \textbf{PR} & \textbf{Attack} &
\textbf{Total Drift (EDDM)} & \textbf{Overlap} \\
\hline

\multicolumn{5}{|c|}{\textbf{Leveraging Bagging Classifier}} \\
\hline
Na\"ive-LB & 0.4 & NI & 24 & 1 \\
\hline
AT-LB & 0.4 & NI & 2 & 1 \\
\hline
AT-LB & 0.6 & NI & 1 & 1 \\
\hline

\multicolumn{5}{|c|}{\textbf{SRP Classifier}} \\
\hline
AT-SRP & 0.2 & NI & 1 & 1 \\
\hline
AT-SRP & 0.8 & NI & 1 & 1 \\
\hline

\multicolumn{5}{|c|}{\textbf{Adaptive Random Forest}} \\
\hline
AT-ARF & 0.8 & NI & 3 & 1 \\
\hline

\end{tabular}
\end{table}

To conduct a more thorough examination of the identified drift-poison overlap, we assessed the model predictions under the NI setups where an EDDM-detected drift aligned with a contaminated sample. Table~\ref{tab:overlap_metrics} presents the PR, accuracy, F1-score, precision, recall, and overlap for these instances. For LB, at a PR of 0.4, the Naïve-LB model achieved an accuracy and recall of 0.607, an F1-score of 0.709, and a precision of 0.936. At the same PR, AT-LB achieved significantly superior accuracy (0.904), recall (0.904), and F1-score (0.921), while preserving a precision of 0.952. At a heightened PR of 0.6, AT-LB demonstrated comparable efficacy, with an accuracy of 0.903 and an F1-score of 0.921.

For the remaining classifiers, AT-SRP attained accuracy, precision, recall, and F1-score metrics of 0.950 with a PR of 0.2. With a PR of 0.8, the accuracy and recall dropped to 0.884, although precision stayed elevated at 0.955 and the F1-score remained at 0.908. The AT-ARF, assessed at a PR of 0.8, attained an accuracy and recall of 0.938, a precision of 0.934, and an F1-score of 0.936. Significantly, every configuration shown in the table exhibits a singular overlap count. 

\begin{figure*}[t]
    \centering
    \includegraphics[width=0.9\linewidth]{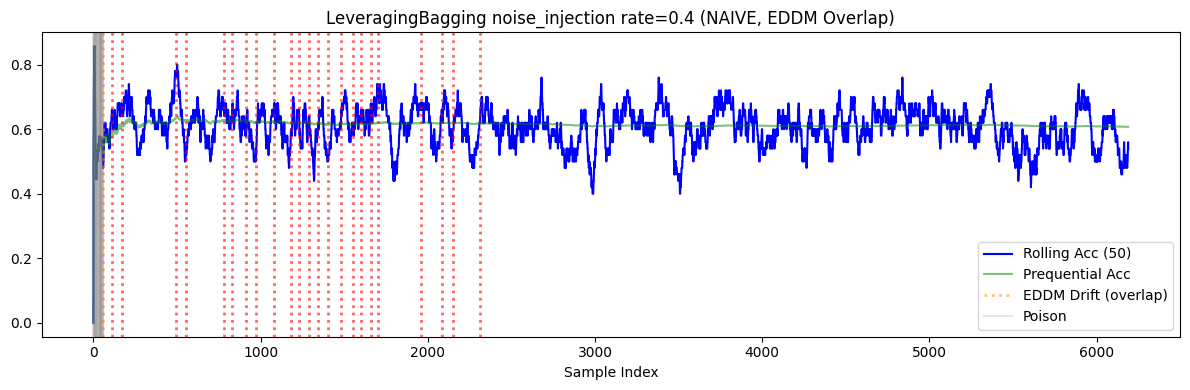}
    \caption{Drift Detection for Naive-LB Model}
    \label{fig:driftdetect1Na}
\end{figure*}

\begin{figure*}[t]
    \centering
    \includegraphics[width=0.9\linewidth]{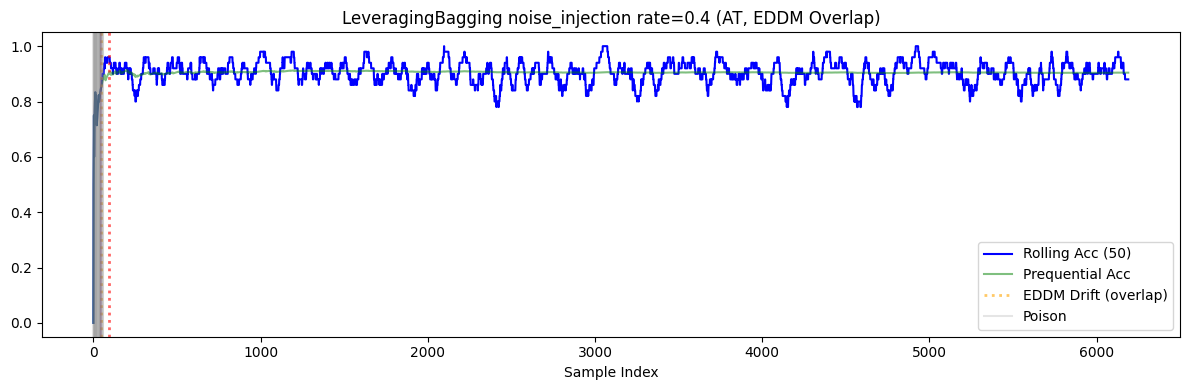}
    \caption{Drift Detection for AT-LB Model}
    \label{fig:driftdetect2AT}
\end{figure*}

Also, whereas these findings delineate model efficacy, particularly for the noise-injection settings when drift-poison overlap was seen, they do not inherently confirm that the coinciding poisoning incident caused the associated performance alterations. Figs. \ref{fig:driftdetect1Na} and \ref{fig:driftdetect2AT} show the rolling and prequential accuracies, EDDM drift overlap of the Naive-LB and AT-LB models for noise injection poison attack at poisoning rate 0.4.

\begin{table}[!t]
\caption{Model Performance for Noise-Injection Cases With Drift-Poison Overlap}
\label{tab:overlap_metrics}
\centering
\begin{tabular}{|l|c|c|c|c|c|c|}
\hline
\textbf{Model} & \textbf{PR} & \textbf{Acc} & \textbf{F1} &
\textbf{Prec} & \textbf{Rec} & \textbf{Overlap} \\
\hline

\multicolumn{7}{|c|}{\textbf{Leveraging Bagging}} \\
\hline
Na\"ive-LB & 0.4 & 0.607 & 0.709 & 0.936 & 0.607 & 1 \\
\hline
AT-LB & 0.4 & 0.904 & 0.921 & 0.952 & 0.904 & 1 \\
\hline
AT-LB & 0.6 & 0.903 & 0.921 & 0.952 & 0.903 & 1 \\
\hline

\multicolumn{7}{|c|}{\textbf{SRP Classifier}} \\
\hline
AT-SRP & 0.2 & 0.950 & 0.950 & 0.950 & 0.950 & 1 \\
\hline
AT-SRP & 0.8 & 0.884 & 0.908 & 0.955 & 0.884 & 1 \\
\hline

\multicolumn{7}{|c|}{\textbf{Adaptive Random Forest}} \\
\hline
AT-ARF & 0.8 & 0.938 & 0.936 & 0.934 & 0.938 & 1 \\
\hline

\end{tabular}
\end{table}

\section{Discussion}
\label{sec:discussion}

At the peak PR of 1.0, resilience differed across various poisoning methods and AT classifiers (Tables \ref{tab:at_label_flip_1} and \ref{tab:at_noise_1}). In the context of label-flip poisoning, SRP attained the superior F1-score of 0.904, whereas Leveraging Bagging recorded the greatest F1-score of 0.933 in scenarios including noise injection. While SRP had marginally superior accuracy compared to Leveraging Bagging during noise injection (0.943 vs 0.936), its F1-score was worse (0.925 versus 0.933). Consequently, with the F1-score as the principal metric, the findings reveal variations in AT model efficacy contingent upon the kind of poisoning attack. 


\begin{table}[!t]
\caption{AT Performance Under Label-Flip Poisoning at Rate 1.0}
\label{tab:at_label_flip_1}
\centering
\begin{tabular}{lcc}
\hline
\textbf{Classifier} & \textbf{Acc.} & \textbf{F1} \\
\hline
Hoeffding Tree          & 0.058 & 0.006 \\
Leveraging Bagging      & 0.543 & 0.661 \\
\textbf{SRPClassifier}  & \textbf{0.906} & \textbf{0.904} \\
Hoeffding Adaptive Tree & 0.703 & 0.778 \\
Adaptive Random Forest  & 0.058 & 0.006 \\
\hline
\end{tabular}
\end{table}

\begin{table}[!t]
\caption{AT Performance Under Noise Injection at Rate 1.0}
\label{tab:at_noise_1}
\centering
\begin{tabular}{lcc}
\hline
\textbf{Classifier} & \textbf{Acc.} & \textbf{F1} \\
\hline
Hoeffding Tree          & 0.721 & 0.793 \\
\textbf{Leveraging Bagging} & \textbf{0.936} & \textbf{0.933} \\
SRPClassifier           & 0.943 & 0.925 \\
Hoeffding Adaptive Tree & 0.803 & 0.848 \\
Adaptive Random Forest  & 0.942 & 0.914 \\
\hline
\end{tabular}
\end{table}

Concept drift denotes the evolving data distribution as time progresses. Identifying this drift is essential for ML models to adjust and sustain optimal performance. Various approaches have been formulated for drift detection; however, EDDM has proved very efficient even for our study herein.  

Essentially, an observed drift should not be construed as conclusive proof of a poisoning assault; instead, these findings delineate the behavior of EDDM in relation to the changing prediction stream amongst the assessed attack circumstances. Our findings suggest that EDDM sometimes generates drift detections that align with some poisoning occurrences; yet, the frequency of such overlapping events is minimal compared to the overall count of detected drifts. 

In the examined experiments, no clear overlap between drift and poison was seen for label-flip assaults, indicating that the identified drift occurrences were not directly associated with the specific samples whose labels were altered. Since the AT defense did not consistently mitigate label-flip poisoning, a more effective AT strategy should be investigated.

\begin{figure}
    \centering
    \includegraphics[width=1.0\linewidth]{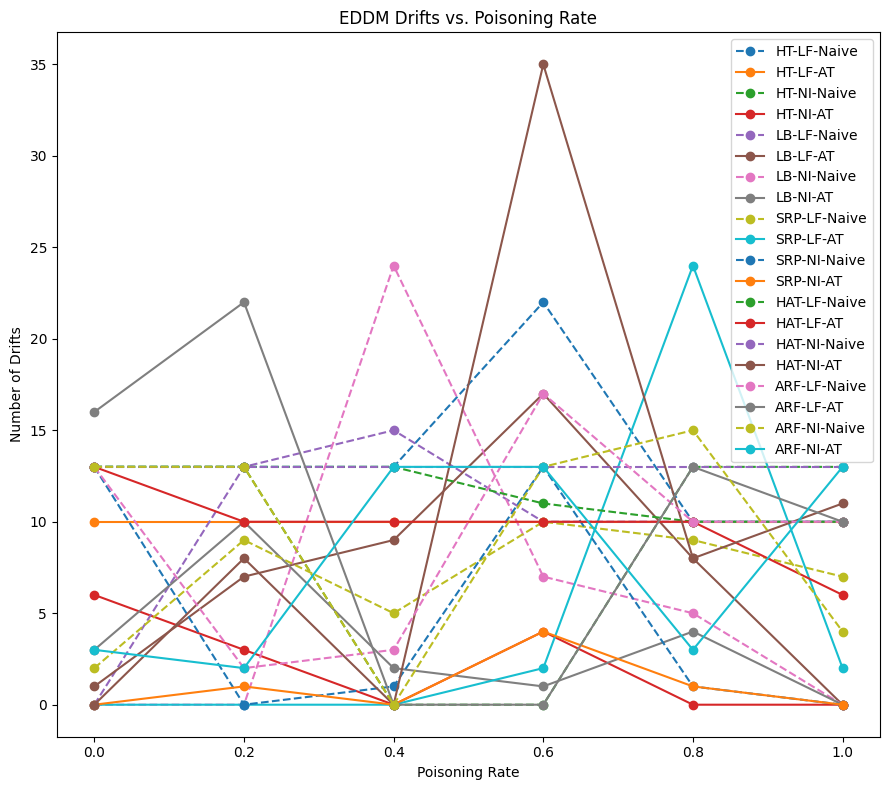}
    \caption{EDDM Detections}
    \label{fig:placeholder}
\end{figure}

\color{black}

\section{Conclusion \& Future Work}
\label{con}

This study examined the impact of poisoning attacks on a streaming AutoML pipeline equipped with an AT defense strategy. It is noteworthy that experiments were conducted at poisoning rates (PR) of 0.0, 0.2, 0.4, 0.6, 0.8, and 1.0. However, for compact presentation, we reported the baseline (0.0), intermediate (0.4), and maximum (1.0) PR. The AT results reveal that SRP classifier is the most robust configuration against label flip poisoning, whereas the LB is the most effective against noise-injection poisoning, evaluated at the peak non-zero poisoning rate (1.0)  using the F1 score. While concept drift detection was done using several methods, EDDM produced more drift detections than the other evaluated detectors. The findings affirm the effectiveness of adversarial training and underscore the need for ongoing model monitoring with adaptive drift detection. Finally, drift detectors were used to detect drift in the naive-LB and AT-LB models. Future work will include statistical significance testing, comparisons across repeated runs, and evaluation on additional IoT datasets.

\section*{Acknowledgment}

This project was made possible in part through the support of the National Cybersecurity Consortium and the Government of Canada (CSIN).

\bibliographystyle{unsrt}
\bibliography{references}

\vspace{12pt}
\end{document}